\documentclass[aps,twocolumn]{revtex4}
\usepackage{graphicx}% Include figure files
\usepackage{dcolumn}% Align table columns on decimal point
\usepackage{bm}% bold math
\usepackage{amsmath,amssymb}
\def\){\right)}
\def\({\left( }
\def\]{\right] }
\def\[{\left[ }

\def\no{\nonumber \\}

\def\be{\begin{equation}}
\def\ee{\end{equation}}
\def\ba{\begin{eqnarray}}
\def\ea{\end{eqnarray}}
\def\no{\nonumber \\}

\def\Z{{\mathbb{Z}}}
\def\C{{\mathbb{C}}}
\def\R{{\mathbb{R}}}

\begin{document}
\setcounter{page}{1}
\title[]{Decays of Four Intersecting Fluxbranes}
\author{Sunggeun  \surname{Lee}}
\email{sglee@photon.khu.ac.kr}
%\affiliation{Department of Physics and Research Institute for Basic Sciences,\\
%Kyung Hee University, Seoul, 130-701, Korea}
\author{Soonkeon  \surname{Nam}}
\email{nam@khu.ac.kr}
\affiliation{Department of Physics and Research Institute for Basic Sciences,\\
Kyung Hee University, Seoul 130-701, Korea}
%\date[]{Received January 5 2004}

\begin{abstract}
We consider decays of four intersecting fluxbranes which are obtained
by considering a higher dimensional Kerr blackhole with four angular
momentum parameters, which is the maximum number of angular momentum parameters
in string/M-theory. As a result of the intersection, we get lower dimensional
fluxbranes. Since generic magnetic fields break all
supersymmetries, the resulting fluxbranes are unstable and will
decay. Just as a single fluxbrane decays into the nucleation of
spherical D6-branes; the intersecting ones decay into the nucleation of lower
dimensional spherical branes. Contrary to a single fluxbrane case, the
decay of four intersecting fluxbranes has additional decay channels.
We also calculate the corresponding Euclidean
action to obtain the decay rates. Although the action cannot be explicitly and simply
written in terms of the magnetic parameters, we can extract some interesting
results by taking various limits of the magnetic parameters.
\end{abstract}

\pacs{04.65.+e,11.25.Uv,11.25.Yb}

\keywords{M-theory, Kaluza-Klein, bubble}

\maketitle

%%%%%%%%%%%%%%%%%%%%%%%%%%%%%%%%%%%%%%%%%%%%%%%%%%%%%%%
\section{Introduction}
%%%%%%%%%%%%%%%%%%%%%%%%%%%%%%%%%%%%%%%%%%%%%%%%%%%%%%%
\setcounter{equation}{0}

Obtaining an exact spectrum in nontrivial backgrounds is quite restricted
in string theory. One example of such a nontrivial background is the Melvin
background\cite{melvin}, which originally appeared as a solution of the
Einstein-Maxwell system.
There the magnetic field, which is gravitating, is aligned along one spatial direction
in four spacetime dimensions.
In string theory, there are two kinds of magnetic fields coming from NS-NS and
R-R gauge
fields and giving NS-NS Melvin background and R-R Melvin background, respectively.
The R-R Melvin background is often
called a fluxbrane. These Melvin backgrounds are related by a U-duality. Since the
nontrivial magnetic fields are gravitating, the spacetime will be curved.
Nevertheless, in the NS-NS Melvin background \cite{russotsey1}, one can calculate
the exact string spectrum.
However, in the fluxbrane background,
the exact string spectrum is not yet available due to the nonperturbative effect of
the dilaton.

Because the Killing spinors do not exist in these Melvin backgrounds
the supersymmetry
is generically broken. In that case, a closed string has tachyons as
a ground state, which is
a signal of instability.
However, if one turns on two or more branes that intersect, there is a
possibility of preserving partial supersymmetry.

As already has been discussed before in Ref.\cite{dowker1}, the unstable
fluxbrane in four
dimensions decays into the nucleation of a pair creation of black holes.
This is the magnetic analog of Schwinger pair
production in an electric field. Later, this idea
was generalized to general dimensional cases \cite{dowker2},
and various decay modes were discussed. One of them was
the creation of charged or uncharged spherical branes,
including Witten's bubble of nothing\cite{bubble}. In addition, intersecting
fluxbranes were discussed, in which the lower-dimensional fluxbranes were
obtained as a result. These lower-dimensional Melvin
backgrounds can be obtained from Kaluza-Kline reduction from higher dimensions
with
a non-trivial identification of the circle. Another way to get a lower dimensional
fluxbrane is to directly choose a metric ansatz and solve the
Einstein equation \cite{gutperle}.
Recently, this was discussed in relation to
M-theory in a general case \cite{russotsey1}. The technique of obtaining magnetic
backgrounds in four dimensions was applied to M-theory\cite{costa}. By calculating
the partition function, type IIA and
type 0A strings were shown to be related by a shift of the magnetic field.
This is an explicit example of the duality between type IIA theory
and type 0A theory\cite{berg}.
From a spin structure argument, one can see that the decays are different
for type IIA and type 0A.
In a type IIA string, the Melvin background decays into the nucleation of a pair
of D6 and anti-D6 branes and
for type 0A string, it decays into a bubble of nothing.
More recently, two intersecting fluxbranes and their decays were
considered in Ref.\cite{brecher}.
By numerical study the Euclidean action was shown to become infinite, and
the fluxbranes were shown not to decay when
the magnetic fields took on particular value. This is consistent
with the existence of supersymmetry.
In the presence of more intersecting branes, one can reduce along Killing vectors
in more
ways than as in the single-brane case. Thus, there are
more decay modes and different kinds of brane creations corresponding to each Killing
vector. Other properties of fluxbranes are discussed in Refs.\cite{empa,brecsaf,figu,
urang,empagut}.

In general, an Fp-brane in $D$-dimension has $ISO(p,1)\times SO(D-p-1)$ symmetry
and a non-zero rank $(D-p-1)$ field strength, a dual field strength or a wedge product
of field strengths tangent to the transverse dimensions. These Fp-branes
couple to a D(p-1)-brane. When the fluxbranes intersect,
the isometry reduces and results in lower-dimensional fluxbrane.
In our work, we consider four intersecting fluxbranes. In particular, we will consider
four intersecting seven-dimensional fluxbranes called F7-branes.
This number of fluxbranes is the
maximum possible number when we consider the positiveness of dimensions of the
created branes. This maximum
number of fluxbranes also comes
if one considers the duality to the NS-NS Melvin background and its orbifold limit:
by taking
appropriate limits, we can obtain orbifolds like $\C^r/\Z_n$, and $r=4$ is the maximum
value for $r$ in 10-dimensions \cite{namsin}.
If we consider five intersecting fluxbranes,
we get a fluxbrane with a negative ($-1$) dimension,
which has no interpretation in string theory.
As can be expected, as we increase the number of intersecting fluxbranes,
there are more ways of reduction corresponding to each Killing vector.
Thus, we can see a number of decay modes as a result.
We can check the existence of the supersymmetry. In an indirect way,
we can construct a partition function for the general Melvin background (we have to
consider the exactly solvable NS-NS Melvin background which is U-dual to a fluxbrane).
Then, this partition function will vanish when the magnetic fields
satisfy a particular relation. As an example, for four magnetic fields, the
partition function \cite{takaya} vanishes when the magnetic fields
satisfy $B_1=\pm B_2\pm B_3\pm B_4$.
When the magnetic fields
do not satisfy this relation, the closed string tachyonic modes are
present \cite{aps}. For the discussion of the rolling string
tachyon, see for example Ref.\cite{wsyi}. Another way is to directly
solve the Killing spinor equation. The existence of the
solution of the Killing spinor equation is known to be consistent with a vanishing
partition function \cite{russotsey2}.

The rest of the paper is as follows: In Section II, we briefly review the Kaluza-Klein
magnetic flux tube background and its decay. In Section III,
we will consider four intersecting fluxbranes and
their decay. With the intersecting fluxbranes, we can consider more Killing
vectors and corresponding decay channels. We also calculate the Euclidean action.
Since we cannot obtain the action as an explicit function of magnetic fields,
it is not easy to analyze the result. However, when we take limits, it is much easier
to analyze it. In Section IV,
we conclude with some discussions.

%%%%%%%%%%%%%%%%%%%%%%%%%%%%%%%%%%%%%%%%%%%%%%%%%%%%%%%%%%%%%%%%%%
\section{Kaluza-Klein magnetic flux tube background and its decay}
%%%%%%%%%%%%%%%%%%%%%%%%%%%%%%%%%%%%%%%%%%%%%%%%%%%%%%%%%%%%%%%%%%
In Ref.\cite{bubble}, Witten pointed out that the original Kaluza-Klein vacuum
${\rm M}_4\times S^1$ is semiclassically unstable and decays into a bubble of nothing.
He started with a five-dimensional Euclidean flat space, where one direction was
periodically
identified on a circle. The fact that this space is unstable can be checked directly
by considering fluctuations around the background, and
exponentially growing modes are found, i.e., the existence of
a (gravitational) instanton (or a bounce) solution.
In five dimensions, the Euclidean Schwarzschild black hole solution acts as the bounce
solution. Therefore,
one can insist that the Kaluza-Klein vacuum decay through an instanton into the
bubble of nothing. One can extend this idea to many other cases that have an instanton
solution.
One trivial extension is to regard any D-dimensional Euclidean Schwarzschild black hole
as the bounce solution.

Now we can generalize Witten's idea to the case with a magnetic field. For that purpose,
we need to consider the Melvin background.
The four dimensional Melvin universe \cite{melvin} can be obtained by dimensional
reduction from five dimensions,
\be
ds^2=-dt^2+dz^2+d\rho^2+\rho^2 d\phi^2+(dx_5)^2, \label{1}
\ee
with nontrivial identification
\be
(t,z,\rho,\phi,x_5)\equiv(t,z,\rho,\phi+2\pi n_1 R B+2\pi n_2,x_5+2\pi n_1 R),
\label{2}
\ee
where $n_1,n_2\in \Z$ and $B$ is the magnetic field.
In this case, since we have to twist both the $\phi$ and the $x_5$
directions, the Killing vector is given by
\be
K={\partial \over {\partial x_5}}+B{\partial \over {\partial \phi}}.
\ee
The four (from $D=5$) reduced dimensions can be read by using the general
Kaluza-Klein ansatz
\ba
ds_D^2&=&{\rm exp}\({4\phi \over {\sqrt{D-2}}}\)\(dx^D+2A_\mu dx^\mu\)^2\no
&+&{\rm exp}\(-{4\phi \over {(D-3)\sqrt{D-2}}}\)g_{E,\mu\nu}dx^\mu dx^\nu,
\ea
from which we can see that the action
\be
S={1\over {16\pi G_D}}\int d^Dx \sqrt{-g_D}R(g_D)
\ee
reduces to
\ba
S&=&{1\over {16\pi G_{D-1}}}\int d^{D-1}x \sqrt{-g}\[R(g_{D-1}) \right. \no
&-&\left. {4\over {D-3}}
(\nabla\phi)^2-{\rm exp}\({-{4\sqrt{D-2}\over {D-3}}}\phi\)F^2\],
\ea
where $2\pi R G_{D-1}=G_D$.
This action contains the dilaton $\phi$ and the gauge field strength $F_{\mu\nu}$,
as well as the metric $g_{\mu\nu}$.
The static Melvin solution to this action was studied in detail in Ref.\cite{dowker1}.
It is described by a dilatonic C-metric or Ernst
metric \cite{ernst}, where two blackholes are
accelerating in
opposite directions in a magnetic background.

In order to get the Melvin metric in four dimensions from five dimensions,
we introduce ${\tilde \phi}=\phi-B x_5$ in Eq. (\ref{1}) in doing the KK reduction
with the identification Eq. (\ref{2}). Reducing along the Killing vector
$\partial /\partial x_5$,  we get the following solutions for the metric and the fields:
\ba
&&ds_4^2=\sqrt{\Lambda}\[-dt^2 +d\rho^2+dz^2\]+
{1\over \sqrt{\Lambda}}\rho^2d{{\tilde \phi}}^2,\no
&&{\rm exp}\(-{4\phi\over \sqrt{3}}\)=\Lambda=1+B^2\rho^2,
~~A_{\tilde\phi}={B\rho^2\over {2\Lambda}}\label{melvin}.
\ea
As we can see, this solution describes a magnetic flux tube which is aligned
in the $z$ direction
and is transverse to the $(\rho,\phi)$ plane. We can identify
it as a flux one brane or F1-brane.
In Ref.\cite{dowker2}, the decay to the bubble of nothing
(obtained by reducing along $\partial/\partial x_5$)
or the creation of black hole pair
accelerating in a magnetic background (obtained by reducing along
$\partial/\partial x_5+
\partial/\partial\phi$) was discussed.
One noticeable point is the fact
that the solution describing the nucleation of two oppositely charged black holes
coming from the Kerr instanton is the Ernst metric solution.
For the creation of a pair of black holes (as can be compared to a KK monopole)
we need a Kerr instanton, which we will discuss below.
We can embed these into string theory or M-theory.

%%%%%%%%%%%%%%%%%%%%%%%%%%%%%%%%%%%%%%%%%%%%%%%%%%%%%%%%%%%%%%%%%%%%%%%%%%%
%\subsection{Melvin background in M-theory}
%%%%%%%%%%%%%%%%%%%%%%%%%%%%%%%%%%%%%%%%%%%%%%%%%%%%%%%%%%%%%%%%%%%%%%%%%%%
{\bf Melvin background in M-theory: A Review}

In the rest of this section, we consider the Kaluza-Klein magnetic flux tube background
in eleven-dimensions. Upon reduction to ten-dimensions, we get the four-dimensional
analog of Melvin background. Let us begin with the 11-dimensional metric in
cylindrical coordinates:
\be
ds^2=-dt^2+dy_mdy^m+dx^2+d\rho^2+\rho^2d\phi^2+dz^2,
\ee
where $m=1,\cdots,6$. Now, let us take the following identifications:
\ba
&&(t,y_m,x,\rho,\phi,z)\no
&&\equiv(t,y_m,x,\rho,\phi+2\pi n_1 RB+2\pi n_2,z+2\pi n_1 R),
\ea
where $n_1$ and $n_2$ are integers. If we reduce along the Killing vector
\be
K={\partial \over {\partial z}}+B{\partial \over {\partial \phi}},
\ee
we get the Melvin background. It is convenient to introduce the coordinate
${\tilde \phi}=\phi-Bz$. Using the relation between the 11-dimensional
M-theory metric and the 10-dimensional string theory metric
\be
ds_{11}^2=e^{-{2\phi \over 3}}ds_{10}^2 +e^{4\phi \over
3}\(dz+2A_\mu dx^\mu\)^2,
\ee
we get
\ba
&&ds_{10}^2=\sqrt{\Lambda}(-dt^2+dy_mdy^m+dx^2+d\rho^2)+{1\over \sqrt{\Lambda}}
\rho^2d{\tilde \phi}^2,\no
&&~~~e^{4\phi\over 3} =\Lambda\equiv 1+\rho^2 B^2,~~~A_{\tilde
\phi}={B\rho^2 \over {2\Lambda}}.
\ea
This is an extension of the F1 brane of Eq. (\ref{melvin}) and is
called a seven-dimensional fluxbrane or F7 brane.
Here, $B$ is calculated from
\be
B^2={1\over 2}F_{\mu\nu}F^{\mu\nu}|_{\rho=0}.
\ee
The condition $|B|\ll1/R$ is necessary in order to have a
reliable perturbative string theory (small $R$) and a Kaluza-Klein ansatz
($\rho\gg R$). If only bosons are considered, the magnetic field $B$ lies
within the range
\be
-{1\over {2R}}<B\leq {1\over {2R}}.
\ee
When fermions are included,
in order not to change the boundary condition (periodic boundary),
$B$ lies within the following range:
\be
-{1\over R}<B\leq {1\over R}.
\ee

Now, let us see the decay of the Kaluza-Klein magnetic background.
As mentioned above, the instanton we are interested in will be
given by the 11-dimensional Kerr black hole \cite{myperry}:
\ba
ds^2&=&dy_mdy^m+dz^2+\sin^2\theta (r^2-\alpha^2)d\phi^2   \no
&-&{\mu \over \rho^2}
(dz +\alpha \sin^2\theta d\phi)^2+{\rho^2 \over {r^2-\alpha^2-\mu}}dr^2\no
&+&\rho^2d\theta^2 +r^2\cos^2\theta d\psi^2,
\ea
where $\rho^2=r^2-\alpha^2\cos^2\theta$. The apparent singularity is at
$r^2=r^2_H\equiv \mu +\alpha^2$. We express the parameters of this background
as
\be
B={\alpha \over \mu},~~~R={1\over \kappa}={\mu \over \sqrt{\mu+\alpha^2} },
\ee
where $\omega\equiv i\Omega=iB$ is a Lorentzian angular velocity
and $\kappa$ is a surface gravity. If we set
$\alpha$ to zero the metric becomes a flat 11-dimensional Euclidean space. We can
dimensionally reduce it
in two ways: either $K=\partial/\partial z+\Omega \partial /\partial \phi$ or
$K'=\partial /\partial z+(B\pm (1/R))\partial/\partial z$, which have zero norm
at $r=r_H$. The reduction can be done by redefining ${\tilde \phi}=\phi-Bz$
or ${\tilde \phi}=\phi-(B\pm(1/R))z$. The latter case is called the shifted instanton.
The conical singularity at $r=r_H$ can be removed if $z$ has
a period of $2\pi R$.

We can estimate the decay rate as
\be
\Gamma\sim e^{-I_E},
\ee
where $I_E$ is the Euclidean action,
\be
I_E=-{1\over {16\pi G_{11}}}\int d^{11}x \sqrt{g}R-{1\over {8\pi G_{11}}}
\int d^{10}x\sqrt{h}({\cal K}-{\cal K}_0).
\ee
Here ${\cal K}$ is the trace of the extrinsic curvature of the boundary,
and ${\cal K}_0$ is
the same quantity for the Melvin background.
For unshifted Kerr instanton, we have
\be
I_E=\({\pi V_6 \over {8G_{10}}}\) {R^2 \over {1-(BR)^2}},
\ee
while for a shifted instanton, we have
\be
I_E=\({\pi V_6 \over {8G_{10}}}\){R^2 \over {1-(1-|B|R)^2}}.
\ee
In the above, $V_6$ is the volume of a unit 6-sphere.

Since the Kerr instanton
has a topology of $\R^2\times S^3$, it admits a single-spin structure. After a
parallel transport around a closed curve at infinity, the spin picks up a phase
of $-e^{\pi R
\alpha \Gamma/\mu}$ for both the shifted and the unshifted instanton.
On the other hand, the magnetic background has a topology of a circle (since ${\rm M}^4
\times S^1$)
and thus there can be two types
of spin structures:
\be
e^{\pi R B\Gamma},~~~{\rm and}~~~-e^{\pi R B \Gamma},
\ee
where
$\Gamma$ is an element of the Lie algebra of Spin(1,10) satisfying
$\Gamma^2=-1$.
If we choose the first spin structure corresponding to type IIA, the
appropriate Kerr instanton is the shifted one, $-e^{\pi
R(B\mp 1/R)\Gamma}=+e^{\pi R B \Gamma}$ ($\alpha/\mu=B\pm (1/R))$, whose
Killing vector is $K'=K\mp{1\over
R}{\partial \over {\partial \phi}}$.
We choose the positive sign for a positive $B$ field while we choose a negative sign for
negative $B$ field.
The second choice picks an unshifted instanton as the decay mode for type 0A,
whose Killing vector is $K={\partial \over {\partial z}}+
B{\partial \over {\partial \phi}}$.
The instanton of type 0A decays to Witten's bubble of nothing, and
that of type IIA decays to pair creation of 6-branes.

%%%%%%%%%%%%%%%%%%%%%%%%%%%%%%%%%%%%%%%%%%%%%%%%%%%%%%%%%%%%%%%%%%
\section{Four intersecting fluxbranes}
%%%%%%%%%%%%%%%%%%%%%%%%%%%%%%%%%%%%%%%%%%%%%%%%%%%%%%%%%%%%%%%%%%
The fixed point sets under the Killing vectors isometry correspond to
branes or decay products after reduction. We will apply the results
of Ref.\cite{dowker2} to our case.

In order to discuss the decay of four intersecting fluxbranes, we have to consider the
11-dimensional Kerr black hole \cite{myperry} with four angular momentum parameters
for each transverse $\R^2$. Before studying that, let us briefly review
intersecting fluxbranes as discussed in Ref.\cite{dowker2}. Consider 11-dimensional metric
\be
ds_{11}^2=\sum_{i=1}^{5}(d\rho_i^2+\rho_i^2d\phi_i^2)+dy^2.
\ee
Since a fluxbrane, for example, F1 in four dimensions, can be considered
as a magnetic field piercing through
two transverse plane, intersecting fluxbranes in higher dimensions can be obtained
by taking two orthogonal planes and turning on a magnetic flux
on each plane. This can be done by the following:
We reduce along the Killing direction with the Killing field
\be
K={\partial \over {\partial y}}+\sum B_i {\partial \over {\partial \phi_i}}.
\ee
If $\tilde{\phi}_i=\phi_i-B_i y$ is introduced, the new Killing field will be
$K'=\partial/\partial y$. We rewrite the metric as
\ba
ds^2 &=&\Lambda \[ dy +\Lambda^{-1} \sum_i B_i \rho_i^2 d\tilde {\phi_i}\]^2
+\sum_i (d\rho_i^2 +\rho_i^2 d\tilde{\phi_i}^2)\no
&-&\Lambda^{-1}\(\sum_j B_j \rho_j^2 d\tilde {\phi_i}^2\)^2,
\ea
where
\be
\Lambda =1+\sum_i B_i^2 \rho_i^2.
\ee
Then, the reduced 10-dimensional metric will be
\ba
ds^2_{10}&=&\Lambda^{1\over 8}\[ -dt^2 +\sum_i(d\rho_i^2 +\rho_i^2
d\tilde{\phi_i}^2)\right.
\no &=&\left.-\Lambda^{-1}\(\sum_j B_j\rho_j^2 d\tilde{\phi}^2\) \]\no
\Lambda &=&e^{-{4\over \sqrt{7}}\phi},~~~A={1\over {2\Lambda}}
\sum_I B_i \rho_i^2 d\tilde{\phi_i}.
\ea
The case of a F7-brane arises when only one direction is chosen.
In this case, the F7-brane in $11$-dimension has $ISO(7,1)\times SO(3)$ symmetry
and a non-zero-rank three field strength, a dual field strength, or a wedge product
of field strengths tangent to the transverse dimensions.
From the above general result, the extension to
four $B_i$ nonzero, which is our main consideration here,
by reducing the Poincare symmetry with the choice of the four $\tilde{\phi}_i$, is easy.
As a result
of turning on more magnetic parameters, the dimensionality of the resulting
fluxbrane is reduced.

Now, let us discuss the decay of four intersecting fluxbranes.
The 11-dimensional metric we are considering with four angular-momentum
parameters is given by \cite{myperry}
\ba
ds_{11}^2&=&dz^2+\sum_{i=1}^4(r^2-\alpha_i^2)(d\mu_i^2+\mu_i^2d\phi_i^2)\no
 &+&r^2(d\mu_5^2+\mu_5^2 d\phi_5^2)+{\Pi F \over {\Pi -\mu
r^2}}dr^2\no
&-&{\mu r^2 \over {\Pi F}}(dz^2 +\sum_{i=1}^4\alpha_i \mu_i^2 d\phi_i)^2,
\ea
where
\ba
\Pi=r^2\prod_{i=1}^4(r^2-\alpha_i^2),~~~~~
F=1+\sum_{i=1}^4 {\alpha_i^2\mu_i^2\over {r^2-\alpha_i^2}},
\ea
and
\be
\mu
=\prod_{i=1}^4(r_H^2-\alpha_i^2),~~
\ee
with the constraint of
\be
\mu_5^2=1-\sum_{i=1}^4\mu_i^2 \label{mu}.
\ee
To avoid a conical singularity at $r=r_H$, we must identify the
coordinates as
\ba
&&z\equiv z+2\pi R,\no
&&\phi_i\equiv \phi_i+2\pi\Omega_i R+2\pi m_i,~~~(i=1,2,3,4),
\ea
where
\be
R={2\mu r_H^2\over {\Pi'(r_H)-2\mu r_H}} =\mu{1\over\Sigma} \label{r}
\ee
and
\ba
\Sigma&=&r_H[4r_H^6-3r_H^4(\alpha_1^2+\alpha_2^2+\alpha_3^2+\alpha_4^2)\no
&+&2r_H^2((\alpha_1^2+\alpha_2^2)(\alpha_3^2+\alpha_4^2)
+\alpha_1^2\alpha_2^2+\alpha_3^2\alpha_4^2))\no
&-&(\alpha_1^2+\alpha_2^2)\alpha_3^2\alpha_4^2-(\alpha_3^2+\alpha_4^2)
\alpha_1^2\alpha_2^2]
\ea
which can be rewritten as
\be
r_H[(r_H^2-\alpha_1^2)(r_H^2-\alpha_2^2)(r_H^2-\alpha_3^2)
+(2,3,4)+(1,3,4)+(1,2,4)],\label{sigma}
\ee
with $(i,j,k)$ meaning $(r_H^2-\alpha_i^2)(r_H^2-\alpha_j^2)
(r_H^2-\alpha_k^2)$ in the brackets.
In the asymptotic limit, the solution goes to flat spacetime with non-trivial
identification. In this region, the solution looks like the Euclidean 11-dimensional
intersecting fluxbrane (F1) solution. This causes us to identify the magnetic
fields as
\be
B_i=\Omega_i~~~{\rm for}~i=1,2,3,4.
\ee

%%%%%%%%%%%%%%%%%%%%%%%%%%%%%%%%%%%%%%%%%%%%%%%%%%%%%%%
\subsection{Spin Structure and Decay of Fluxbranes}
%%%%%%%%%%%%%%%%%%%%%%%%%%%%%%%%%%%%%%%%%%%%%%%%%%%%%%%
%%%%%%%%%%%%%%%%%%%%%%%%%%%%%%%%%%%%%%%%%%%%%%%%%%%%%%%
%\subsubsection{Spin structure}
As we saw in the previous section, the Killing vector is important
if we want to consider the decays of the fluxbrane.
It determines whether the result is
type IIA or type 0A
and the types of branes created. The types of nucleation
depend on the fixed point sets of the Killing vector.

Now we briefly see what the fixed point sets are and how they are related to
the decay product by looking into the examples
summarized in Ref.\cite{dowker2}
and references therein. Consider the isometry generated by the Killing vector $K$.
For the associated Killing field $f_a$, we define
\be
f_{ab}\equiv f_{a;b}.
\ee
At a fixed point, $f_a=0$. In addition, let the kernel of $f_{ab}$ have dimension $p$.
The vectors in the kernel are directions in the tangent space at the fixed point
and are
invariant under the action of the symmetry. The $p$-dimensional subspace of fixed
points
is what will be nucleated after decay of fluxbranes. In four dimensions, $p=0$ is
called a `nut', and $p=2$ is called a `bolt'. The simple example is a five-dimensional
Euclidean Schwarzschild solution:
\ba
ds^2&=&dz^2+ \( 1-{2m\over r}\) d\tau^2 + \(1-{2m\over r}\)^{-1}dr^2\no
&+&r^2 (d\theta^2 + \sin^2\theta d\phi^2),
\ea
where $\tau$ has period $2\pi R$ with $R=4m$. If we reduce along the Killing vector
$\partial/\partial\tau$, we find a minimum of two-sphere or bolt as the
fixed point set of that circle action. Consider now the alternative Killing vector
$\partial/\partial\tau+(1/R)\partial/\partial\phi$. This has fixed points at the north
and the south poles, being nut$(\theta=0)$ and anti-nut$(\theta=\pi)$, respectively.
We apply this idea to following decays:

In our case of four intersecting fluxbranes,
various kinds of Killing vectors and the corresponding decays can be considered.
Now let us consider them one by one.

%%%%%%%%%%%%%%%%%%%%%%%%%%%%%%%%%%%%%%%%%%%%%%%%%%%%%%%
%\subsubsection{decay}
\begin{itemize}
\item
Reduction along the Killing vector
\be
K_1={\partial \over
{\partial z}}+\sum_{i=1}^4 \Omega_i{\partial \over {\partial \phi_i}}.
\ee
The result will give $B_i=\Omega_i$ ($i=1,\cdots 4$) and the fixed-point
set is a 9-sphere (entire horizon), which is just
Witten's bubble of nothing. From a consideration of the spin structure,
this results in a type-0A theory.
\item
Reduction along the Killing vector
\be
K_2=K_1\mp{1\over R}{\partial \over
{\partial \phi_i}}.
\ee
The result will give magnetic fields
$B_i=\Omega_i\mp{1\over R}$ and $B_{j\neq
i}=\Omega_{j\neq i}$. There are four ($_4{\rm C}_1$) types of decays,
and the fixed point set is a charged spherical 6-brane. From a consideration
of the spin
structure, this results in a type IIA theory.
\item
Reduction along the Killing vector
\be
K_3=K_1\mp{1\over R}{\partial \over {\partial \phi_i}}\mp{1\over
R}{\partial \over {\partial \phi_j}}, ~~i\neq j.
\ee
There are six types
($_4{\rm C}_2$) of decays. The result will give magnetic fields
$B_i=\Omega_i\mp{1\over R}$ and $B_j=\Omega_j\mp{1\over R}$, and
the fixed point sets are uncharged spherical 4-branes. From a consideration of
the spin
structure, this results in type 0A theory.
\item
Reduction along the Killing vector
\be
K_4=K_1\mp
{1\over R}{\partial \over {\partial \phi_i}}\mp {1\over
R}{\partial \over {\partial \phi_j}}\mp{1\over R}{\partial \over
{\partial \phi_k}},~~i\neq j\neq k.
\ee
There are four ($_4{\rm C}_3$) types
of decays. The result will give magnetic fields
$B_{i,j,k}=\Omega_{i,j,k}\mp{1\over R}$, and the fixed point set
is a uncharged spherical 2-brane. From the spin structure, this results in
a type IIA theory.
\item
Reduction along the Killing vector
\be
K_5=K_1\mp {1\over
R}\sum_{i=1}^4{\partial \over {\partial \phi_i}}.
\ee
There is a single
($_4{\rm C}_4$) type of decay. The result will give magnetic fields
$B_i=\Omega_i\mp{1\over R}$, and the fixed point set is an
uncharged 0-brane. From a consideration of the spin structure,
this results in a type 0A theory.
\end{itemize}
The first, the second, and the fifth cases are of the type that were considered
in Ref.\cite{brecher}.
The third and the fourth cases are new ones, which are due
to the intersection of more than two fluxbranes.

%%%%%%%%%%%%%%%%%%%%%%%%%%%%%%%%%%%%%%%%%%%%%%%%%%%%%%%
\subsection{Calculation of the Action}
%%%%%%%%%%%%%%%%%%%%%%%%%%%%%%%%%%%%%%%%%%%%%%%%%%%%%%%
The decay rate of the fluxbranes is equal to the creation rate of the spherical
D-branes and is given by
$e^{-I_E}$. To see this explicitly, we have
to calculate the Euclidean action. For this calculation, we apply the method
introduced in Ref.\cite{dowker2}.
The instanton we are considering is Ricci flat, and the
contribution to the action is zero. Thus the only non-vanishing action comes
from the boundary term
\be
I_{E,11}=-{1\over {8\pi
G_{11}}}\int_{r\to \infty}d^{10}x\sqrt{h}({\cal K}-{\cal K}_0),
\ee
where $h$ is
the determinant of the induced metric on a constant $r$ surface,
${\cal K}$ is the extrinsic curvature of this metric, and ${\cal K}_0$ is the
extrinsic curvature of the reference background (in our case
$\mu=0$).

The induced metric $r=0$ can be written as follows:
\ba
&&ds_{10}^2=dz^2
+\sum_{i=1}^{4}(r^2-\alpha_i^2)(d\mu_i^2+\mu_i^2d\phi_i^2)
\no
&&+r^2(d\mu_5^2+\mu_5^2d\phi_5^2)-{\mu r^2\over {\Pi F}}
(dz+\sum_{i=1}^4 \alpha_i \mu_i^2d\phi_i)^2,
\ea
where we have to put
\be
\mu_5^2=1-\sum_{i=1}^4\mu_i^2,~~~d\mu_5^2
=(\sum_{i=1}^4 \mu_id\mu_i)^2
\ee
in the metric.

We will use the relation \cite{dowker2}
\be
\sqrt{h} {\cal K}={\hat n}\sqrt{h},
\ee
where the unit normal vector ${\hat n}$ is given by
\be
{\hat n}=\sqrt{
{{\Pi -\mu r^2} \over {\Pi F}} } {\partial \over {\partial r}}.
\ee
From the given metric, we can get the determinant $\sqrt{h}$, which is given as follows:
\be
\sqrt{h}=(\prod_{i=1}^4 \mu_i) \sqrt{ (r^8+D r^6 -Cr^4+Br^2-A)(\Pi-\mu r^2)},
\ee
where
\ba
A&=&\(\prod_{i=1}^4 \alpha_i^2\)\((\sum_{j=1}^4 \mu_i^2) -1\) ,\no
B&=&\(\prod_{i=1}^4 \alpha_i^2\) \sum_{j=1}^4 { \( {(\sum_{k=1}^4 \mu_k^2) -\mu_j^2 -1}
 \over \alpha_j^2 \)}, \no
C&=&\sum_{i,j=1,i<j}^4\alpha_i^2\alpha_j^2(\mu_i^2+\mu_j^2-1),\no
D&=&\sum_{i=1}^4\alpha_i^2(\mu_i^2-1).
\ea
Notice that for the case of two intersecting fluxbranes ($\alpha_3=\alpha_4=0$),
our result reduces to the one in Ref.\cite{brecher}, i.e.,
\ba
A&=&0,~
B=\alpha_1,~
C=\alpha_1^2\alpha_2^2(\mu_1^2+\mu_2^2-1),\no
D&=&\alpha_1^2(\mu_1^2-1)+ \alpha_2^2(\mu_2^2-1).
\ea

We can evaluate $\sqrt{h}{\cal K}={\hat n}\sqrt{h}$ as
\ba
&&{\hat n}\sqrt{h}
={\mu_1\mu_2\mu_3\mu_4 \over 2}\no
&&~~\times [{ (\Pi-\mu
 r^2)(8r^7+6Dr^5-4Cr^3+2Br)\over { \sqrt{r^8+Dr^6-Cr^4+Br^2-A}
  }}   \no
&&~~+{(r^8+Dr^6-Cr^4+Br^2-A)(\Pi'-2\mu r) \over {
 \sqrt{r^8+Dr^6-Cr^4+Br^2-A} } }].
\ea
Since we can take the limits
\be \lim_{r\to \infty}
{\sqrt{h} \over {\sqrt{h}|_{\mu=0}}}=1-{1\over 2}{\mu \over
r^8},~~~~~{\cal K}={{\hat n}\sqrt{h} \over \sqrt{h} },
\ee
we can obtain the integrand in the boundary term as
\ba
&&\lim_{r\to \infty} \sqrt{h}({\cal K}-{\cal K}_0)\no
&=&\lim_{r\to \infty}
\({\hat n}\sqrt{h}-{\hat n}\sqrt{h}|_{\mu=0}+{1\over 2} {\mu \over r^8}
{\hat n}\sqrt{h}|_{\mu=0}\)\no
&=&\lim_{r\to \infty}\( {\partial \over {\partial
\mu} } {\hat n}\sqrt{h}+{1\over 2} {1\over r^8} {\hat n}\sqrt{h}|_{\mu=0} \)\mu.
\no
\ea
This becomes
\begin{widetext}
\ba
%&&\lim_{r\to \infty} \sqrt{h}({\cal K}-{\cal K}_0)=
\lim_{r\to \infty}
{\mu_1\mu_2\mu_3\mu_4 \over 2}\mu
%\no
%&&~~~~~\times
\[ {-r^2(8r^7+6Dr^5-4Cr^3+2Br)+(r^r+Dr^6-Cr^4+Br^2-A)(-2r) \over{
\sqrt{ \Pi F (r^8+Dr^6-Cr^4+Br^2-A) } } }\right.\no
%&&~~~~~
\left.+{1\over 2} {1\over
r^8} { \Pi(8r^7+6Dr^5-4Cr^3+2Br)+(r^8+Dr^6-Cr^4+Br^2-A)\Pi' \over
{ \sqrt{ \Pi F(r^8+Dr^6-Cr^4+Br^2-A)} } } \].
\ea
\end{widetext}

Surprisingly, the final result is simplified to
\be
\lim_{r\to \infty} \sqrt{h}({\cal K}-{\cal K}_0)=-{\mu_1\mu_2\mu_3\mu_4 \over 2}\mu.
\ee
This final result looks exactly like the result obtained in Ref. \cite{brecher} but it
is, in fact, different in that our $\mu$ contains four magnetic parameters instead of two.
Our result is the most general one in that in ten-dimensional string theory
or eleven dimensional M-theory, the upper limit on the number of intersecting
fluxbrane is
four;furthermore, by turning off the magnetic parameters, we get
results that reduce to lower-dimensional,
known decays of two intersecting fluxbranes or a single fluxbrane.

Now, we are ready to
calculate the action. After putting all these into the action, we
get
\be
I_{10,E}={V_9\over {16\pi G_{10}}}\mu,
\ee
where we define
\be
V_9=\int^{2\pi}_0 d\phi_1\cdots d\phi_5\int d\mu_1\cdots d\mu_4 ~\mu_1\cdots \mu_4
\ee
as the
volume of a unit 9-sphere and we have used $G_{10}=2\pi R G_9$.
\begin{table*}[t]
\caption{Divergent points of the Euclidean action $I_{E,11}$ for each Killing vectors}
%\begin{ruledtabular}
\begin{tabular}{|c|c|}
  \hline
  % after \\: \hline or \cline{col1-col2} \cline{col3-col4} ...
  Killing Vector & Divergent points of $I_{11}^E$ $(\Omega_1R_1,\Omega_2R_2,\Omega_3R_3,\Omega_4R_4)$
  \\
  \hline
  $K_1$ & $(\pm 1,0,0,0)$\ \qquad $(0,\pm 1,0,0)$ \ \qquad $(0,0,\pm 1,0)$ \ \qquad $(0,0,0,\pm 1)$  \\
  $K_2$ & $\ \ \ (0,0,0,0)$\ \qquad $(\pm 1,\pm 1,0,0)$\ \quad $(\pm 1,0,\pm 1,0)$\ \qquad $(\pm 1,0,0,\pm 1)$ \\
  $K_3$ & $(0,\pm 1,0,0)$\quad \ \ \ $(\pm 1,0,0,0)$\quad \ \ \ $(\pm 1,\pm 1,\pm 1,0)$\quad \ $(\pm 1 ,\pm 1,0,\pm 1)$ \\
  $K_4$ & $(0,\pm 1 ,\pm 1,0)$\quad \ $(\pm 1, 0,\pm 1,0)$\quad \ $(\pm 1,\pm 1,0,0)$\quad \ $(\pm 1,\pm 1,\pm 1, \pm 1)$ \\
  $K_5$ & $(0,\pm 1,\pm 1,\pm 1)$\ \ $(\pm 1,0,\pm 1,\pm 1)$\ \ $(\pm 1,\pm 1,0,\pm 1)$\ \ $(\pm 1,\pm 1,\pm 1,0)$ \\
  \hline
\end{tabular}
%\end{ruledtabular}
\end{table*}
Now let us analyze the action by taking particular limits.
As in Eq. (\ref{mu}), $\alpha_i'$s, $R$ and $\mu$ have complicated relations
and it is not easy to
directly see the behavior of the action just as for the single-brane
case. It is not possible
to express the action as magnetic fields as before. Hence, in order to read some
physics we
discuss some particular limits. As done in  Ref. \cite{costa} or \cite{brecher},
if one can figure out the behavior or diagram (say for shifted or unshifted) of the
action, one can draw or
figure out the action by shifting the diagram or the values of the action
corresponding to each Killing vector. In our case, we start first with
a Killing vector without shifting and continue to other Killing vectors
with increasing number of shifts.

First, consider the case without a shifting in the magnetic parameters, say the
unshifted instanton.
Note that the angular velocities $\Omega_i$'s are given by
\ba
\Omega_i&=&{\alpha_i \over {r_H^2-\alpha_i^2}}\no
&=&{\alpha_i  (r_H^2-\alpha_j^2)
(r_H^2-\alpha_k^2)(r_H^2-\alpha_l^2) \over \mu},\no &&{\rm where}~i\neq j\neq k\neq l.
\ea
Multiplying Eq.(\ref{r}) by $R$ gives
\ba
\Omega_i R&=& {\alpha_i \mu \over {(r_H^2-\alpha_i^2)\Sigma}} \no
&=&{\alpha_i  (r_H^2-\alpha_j^2)
(r_H^2-\alpha_k^2)(r_H^2-\alpha_l^2) \over \Sigma},\no
&&{\rm where}~i\neq j\neq k\neq l.
\ea
Let us take the limits
\be
\alpha_i\to \pm \infty,~~~~~\alpha_{j\neq i}={\rm constant}\ll\alpha_i
\ee
In this limit we have to take $r_H\to |\alpha_i|$ to keep $R$ fixed
(in this way only one
of the terms survives in Eq.(\ref{sigma}) among the four terms, and this
cancels the denominator), and in order to keep
the radius ($R$) fixed,  take $\mu\to R|\alpha_i|^7$.

For example, take $i=1$, that is, consider $\alpha_1\to \pm \infty$. Then
the other $\alpha$'s are too small to be negligible.
To keep $R$ fixed, take $r_H\to |\alpha_1|$
and $\mu \to R|\alpha_1|^7$. Then, three terms in Eq. (\ref{sigma}) vanish, but
the $(2,3,4)$ term does not. Therefore, $\Omega_1 R$ becomes $\pm 1$ in the above limit.
This critical point
is one of several critical points in which supersymmetry restores and generalizes
the cases for only one magnetic
parameter \cite{dowker2,costa} or two \cite{brecher}, depending
the number of fluxbranes.
We can follow the same procedure
for the other $\alpha$'s. In this way, we find that the action diverges due to
dependence of $\mu$.

In summary, with the help of Eq. (\ref{sigma}), by taking the limit
$\alpha_i \to \infty$, we arrive at
\be
\Omega_i R\to \pm 1,~~~~\Omega_{j\neq i}\to 0 ~~{\rm and}~~\mu\to R|\alpha_i|^7 .
\ee
Thus, in this limit, the Euclidean action $I_E$ goes to infinity so that the decay rate
vanishes, and no decay will happen. Since the parameter space spanned by the $B_iR$
is 4-dimensional, it is difficult to plot the values of $I_E$ as a function of
$B_iR$'s. However, we can easily see that
the value of the action diverges at the following points:
$(\pm 1,0,0,0)$, $(0,\pm 1,0,0)$, $(0,0,\pm 1 ,0)$, and
$(0,0,0,\pm 1)$, where $\pm 1$ is a value of $B_i R$. The value of $I_E$ decreases
smoothly off the divergent peaks. This result corresponds
to a type 0A theory, and there are eight points at which such a divergence occurs.
Now, let us see the next decay mode with one shifted direction.

In this case,
it is enough to shift the magnetic parameters $\Omega_i R\to 1\pm\Omega_i R$
for each direction. The maximum point
where the action diverges is now shifted. For instance, if we reduce
along the $1$-direction, then
the above points are shifted to $(0,0,0,0)$, $(\pm 1,\pm 1,0,0)$,
$(\pm 1,0,\pm 1,0)$, and $(\pm 1,0,0,\pm 1)$. This results in a type IIA theory.
When we act further, one more shifting, we get the result for the third decay
mode: For the reduction along the 1 and the 2 directions, the result we have peaks at the
points $(0,\pm 1,0,0)$, $(\pm 1 ,0,0,0)$, $(\pm 1,\pm 1,\pm 1,0)$,
and $(\pm 1,\pm 1,0,\pm 1)$,
corresponding to a type 0A theory.
For the reduction along the directions 1, 2, and 3 we have
$(0,\pm 1,\pm 1,0)$, $(\pm 1,0,\pm 1,0)$, and $(\pm 1,\pm 1,0,0)$,
$(\pm 1,\pm 1,\pm 1,\pm 1)$, corresponding to a type IIA theory.
For the reduction along all directions we have
$(0,\pm 1,\pm 1,\pm 1)$, $(\pm 1,0,\pm 1,\pm 1)$, $(\pm 1, \pm 1,0,\pm 1)$, and
$(\pm 1,\pm 1,\pm 1,0)$, corresponding to a type 0A theory.
We summarize this simple example in Table 1.
(Remember that for Killing vector $K_1$ there is no shift in
the parameters $\Omega_i R_i$. For other
Killing vectors, $\Omega_i R_i$ has to be shifted to $1\pm \Omega_iR_i$
for corresponding directions):

So far, we considered
the diverging points by taking particular examples.
The interesting case is when
the reduction is done to an odd number of directions. These correspond
to type IIA theories. When three directions are chosen, the maximum
diverging action occurs when $B_1=\pm B_2=\pm B_3=\pm B_4$.
This can also be represented by $B_1=\pm B_2$ and $B_3=\pm B_4$ or
$B_1=\pm B_2 \pm B_3 \pm B_4$. Of course
this result contains the case of one reduced direction (second case
of the above decay) which results in
$B_i=\pm B_j(B_{{\rm others}}=0)$.
Our result is consistent with the result
of the Killing spinor consideration\cite{russotsey2}.

%%%%%%%%%%%%%%%%%%%%%%%%%%%%%%%%%%%%%%%%%%%%%%%%%%%%%%%
\section{Conclusion and discussion}
%%%%%%%%%%%%%%%%%%%%%%%%%%%%%%%%%%%%%%%%%%%%%%%%%%%%%%%
We have considered four intersecting fluxbranes and investigated the decay
modes by looking into the fixed point sets of the Killing vectors and the spin structures.
There are more decay
channels because there are more Killing vectors in the four-intersecting-fluxbrane case
compared to the case of a single or two
fluxbrane decay\cite{brecher}. As a result, we obtained various kinds of branes,
including the ones obtained in Ref. \cite{brecher}. The new ones are spherical
4- and 2-branes and a 0-brane.

Because of the complicated relation between the magnetic
field parameters and the parameters
appearing in the action, we do not have an explicit relation of the action
in terms of the magnetic field parameters.
Since the decay rate of the fluxbranes should be equal to
the creation rate of the branes, we can, in principle, calculate
the action by calculating the DBI action in a type IIA case, and which will
give the result for small magnetic fields. The calculation
for the single-fluxbrane case was done in Ref.\cite{costa}. Extending such
a calculation to an intersecting case would be difficult.

Formerly, the Kerr instanton was considered to get a pair creation
of black holes in a Melvin background. In our case, we obtained a spherical
brane, which is a fixed point set, as a result of decays and
obtained a finite action that gave a finite decay rate,
except for some diverging values.
Note that in the case of F7-brane decay, a spherical 6-brane is created and
can be interpreted as a D6-brane.
However, in considering intersecting
fluxbranes our procedure does not generate
RR gauge fields that couple to some lower D-branes.
It will be interesting to see their stability, i.e., to check whether the created
spherical branes will expand, be stable, or shrink. As in Ref.\cite{aharony}
one can also consider a time-dependent background, such as a de Sitter
background, as a result of decay from the Kerr instanton.
One can further consider some other gravitational
instantons and their decays.
Other interesting cases are a consideration of the
electric-magnetic duality of the same theory.
In the current paper we considered and obtained magnetically charged cases. By duality,
one can obtain electrically charged branes and the
construction of fluxbranes that couple ${\rm D}p-{\bar {\rm D}p}$. It is well understood
that the F7-brane plays a role
in stabilizing the ${\rm D}6-{\bar {\rm D}6}$.

\begin{acknowledgments}
This work is supported by grant number R01-2003-000-10391-0 from the
Basic Research Program of the Korea Science and Engineering Foundation
and also by the Science Research Centers (SRC) program of the Korea
Science and Engineering Foundation (KOSEF) through the Center for Quantum
SpaceTime (CQUeST) of Sogang University (R11-2005-21). We would like to thank
hospitality of Asia Pacific Center for Theoretical Physics (APCTP)
where part of the work has been done.
\end{acknowledgments}

\end{document}